# Geometry-calibrated equilibrium sensing for inverse design of nonlocal topological photonic lattices

*Fatemeh Davoodi*[1,2,*] and *Jeffrey McCord*[1,2]

[1]*Nanoscale Magnetic Materials, Institute of Materials Science, Kiel University, 24143, Kiel, Germany*

[2]*Kiel Nano, Surface and Interface Science KiNSIS, Christian Albrechts University, Kiel, Germany*

[*]*fda@tf.uni-kiel.de*

Topological photonic lattices are commonly designed using short-range Hamiltonians, yet realistic nanophotonic structures are governed by geometry- and wavelength-dependent long-range electromagnetic interactions. Here we introduce a geometry-calibrated quantum equilibrium-propagation framework for inference and inverse design in finite nonlocal plasmonic Su-Schrieffer-Heeger lattices. A two-qubit equilibrium sensor is trained in effective-coupling space to distinguish boundary-localized from trivial finite-lattice responses. Because labels inherited from the nearest-neighbor SSH model become unreliable in the presence of nonlocal hopping, each sample is independently relabeled using nonlocal winding numbers and a finite-gap criterion. On this physics-verified evaluation set, the sensor achieves 99.8% sensitivity and 98.1% specificity. A physics-gated robustness score then ranks verified configurations by boundary response, response contrast, gap stability and compatibility with the selected nonlocality regime. Full-wave extinction spectra of isolated and paired gold nanoparticles establish a geometry-to-coupling calibration linking particle size and separation to wavelength-resolved pair couplings. Projecting this calibration onto the verified coupling landscape identifies a finite plasmonic geometry supporting spectrally distinct corner- and edge-dominated responses at 546 and 642 nm, with sector-to-bulk intensity contrasts of $2.4 \times 10^4$ and $1.5 \times 10^3$, respectively. The framework links realistic electromagnetic geometry to nonlocal topological design, moving beyond nearest-neighbor design rules with physics-verified, geometry-resolved inference.

Topological photonics uses lattice geometry to control optical localization, transport and confinement in photonic crystals, coupled resonators, waveguide arrays and plasmonic lattices.[1-8] Much of this methodology derives from short-range tight-binding models with analytically tractable phase boundaries. The Su-Schrieffer-Heeger (SSH) model provides the canonical example: alternating intracell and intercell couplings divide a one-dimensional lattice into trivial and topological sectors, while higher-dimensional SSH-type systems can support edge- and corner-localized states.[9-17] These models connect lattice geometry, coupling hierarchy and boundary response. Plasmonic lattices and interfaces further support directional excitation and collective topological dynamics.[18-19] Real nanophotonic structures, however, rarely satisfy the nearest-neighbor approximation. Their electromagnetic fields extend across several lattice sites, producing interactions that depend on separation, particle dimensions, wavelength, polarization, material dispersion and the dielectric environment. Long-range radiative, dipolar and photon-mediated interactions can reconstruct photonic band structures and modify the energy, localization and robustness of boundary states.[20-24] In plasmonic lattices, such coupling is not a small correction, but the mechanism through which spatially separated elements interact. This creates a fundamental design conflict. Strong electromagnetic coupling is required to form collective photonic states, yet the same interaction can deform the phase boundaries predicted by ideal SSH models. Nonlocal coupling may shift gap closings, hybridize boundary and bulk modes, and redistribute edge or corner weight across a finite lattice. A geometry satisfying the nearest-neighbor dimerization condition can therefore lose its intended topological classification or exhibit a weak boundary response once longer-range interactions, finite-size effects and spectral dispersion are included. Experiments have similarly shown that a nontrivial bulk classification does not by itself guarantee immunity from backscattering or a strong observable boundary response.[25-27] The inverse problem is also challenging because particle size, wavelength and lattice gaps generally modify several effective couplings simultaneously, preventing topology from being inferred directly from geometry.

Realistic topological design must therefore establish both whether a candidate lies in a physically valid topological sector and whether its boundary response survives the interactions that generate the coupling network. Existing approaches address parts of this problem but do not provide a complete geometry-to-topology route. Bulk invariants classify phases when the Bloch Hamiltonian and its protecting symmetries are known, but do not directly quantify localization in a finite device. This distinction is especially important in higher-order and finite systems, where spectral existence, spatial localization and experimental accessibility are separate questions.[11-17,25] Full-wave simulations include geometry, dispersion, loss and radiative coupling, but exhaustive searches over particle size, gap and wavelength are computationally demanding. Data-driven surrogate and inverse-design methods can accelerate such searches,[32-35] but their outputs do not necessarily preserve the distinction between topological class, finite-system boundary response and design robustness. In nonlocal systems, a high classifier score should not override an invalid winding number or a vanishing spectral gap. A practical framework must therefore combine topology verification, rapid evaluation of boundary response and electromagnetic calibration of the effective Hamiltonian.

Quantum equilibrium propagation provides a Hamiltonian-native basis for this task.[36-38] Derived from Onsager reciprocity, QEP converts an output error into the collective equilibrium response of trainable Hamiltonian terms. Because its inputs, trainable variables and outputs remain encoded in the Hamiltonian, QEP provides a compact sensor of how nonlocal coupling redistributes spectral weight among corner, edge and bulk regions. During inference, the sensor maps the coupling configuration onto a finite-system boundary-response score without post-processing the complete eigenspectrum. This output does not replace a topological invariant, but provides a continuous measure of boundary response within a sector independently verified by winding and gap conditions. Here we develop a geometry-calibrated QEP framework for inference and inverse design in finite nonlocal SSH plasmonic lattices (Fig. 1). The sensor learns edge- and corner-sensitive responses in effective-coupling space, while nonlocal winding numbers and a finite-gap criterion independently verify the physical class. A topology-gated robustness score ranks verified configurations, and full-wave spectra of isolated and paired gold nanoparticles provide the calibration required to project them into physical geometry. The framework thereby connects realistic electromagnetic geometry to physically verified nonlocal topological design.

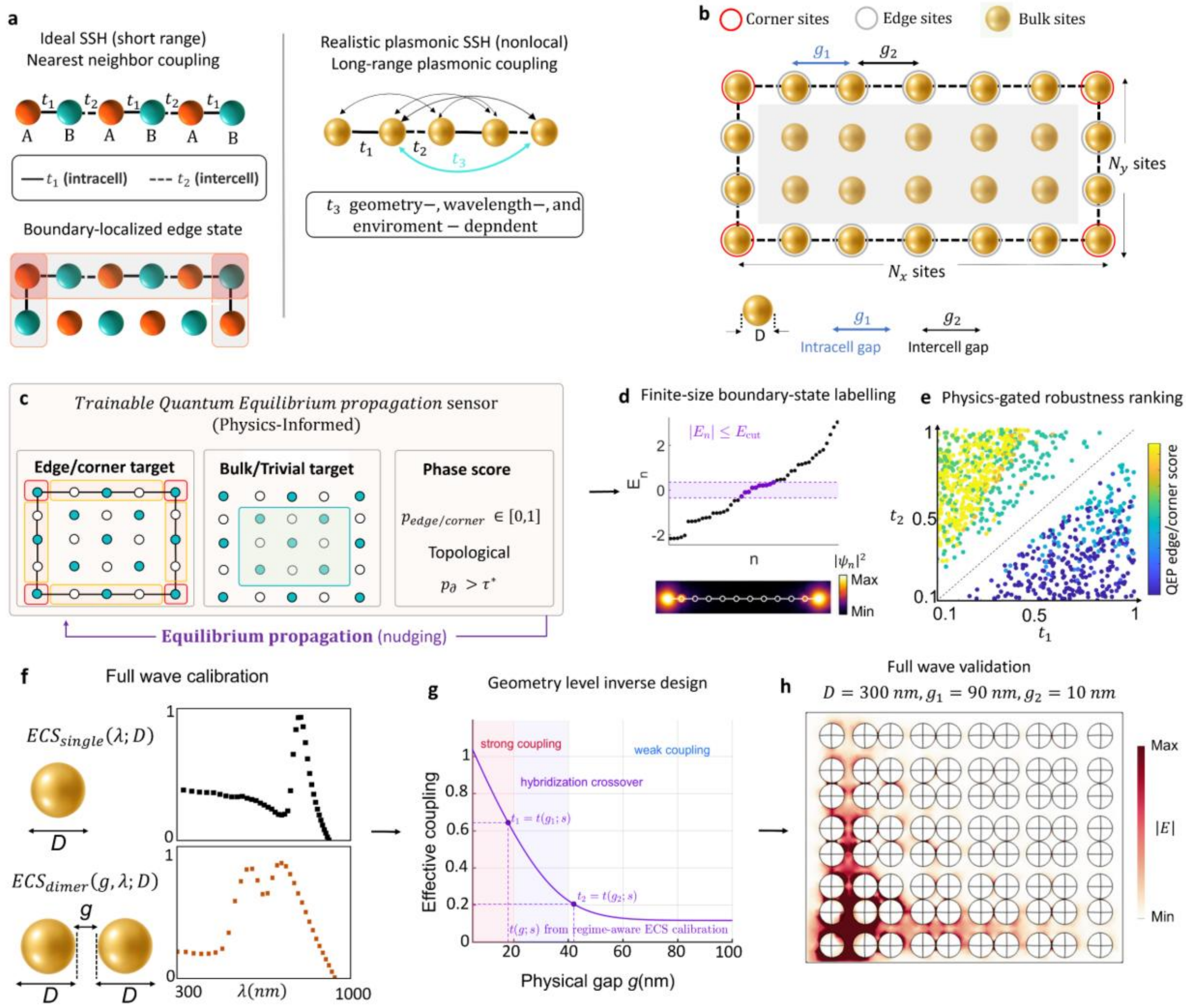

**Fig. 1 | Geometry-calibrated equilibrium sensing for nonlocal topological design. a**, Conceptual comparison between an ideal short-range SSH lattice and a realistic nonlocal plasmonic SSH lattice. In the ideal model, the response is governed by the nearest-neighbor intracell and intercell couplings $t_1$ and $t_2$. In the plasmonic system, the effective longer-range coupling $t_3$ depends on particle geometry, wavelength and local electromagnetic environment, deforming the nearest-neighbor SSH phase structure. **b**, Finite two-dimensional plasmonic SSH lattice used for boundary sensing and inverse design. The structure is partitioned into corner, edge and bulk sectors. The particle diameter is denoted by $D$, while $g_1$, $g_2$ define the intracell and intercell gaps, respectively. **c**, Trainable quantum equilibrium-propagation sensor coupled to the effective nonlocal SSH lattice. The sensor learns to distinguish edge- or corner-localized finite-system responses from bulk or trivial responses through equilibrium output probabilities, with the output mismatch introduced through finite nudging during training. **d**, Finite-size boundary-state labelling used to construct the initial training targets. Candidate states within the near-midgap energy window $|E_n| \leq E_{cut}$ are evaluated according to their spatial localization. **e**, Physics-gated robustness ranking in effective-coupling space. The colour scale shows the QEP edge/corner response score, while the final design ranking additionally incorporates output contrast, gap stability and compatibility with the selected nonlocality regime after topological verification. **f**, Full-wave calibration based on the extinction spectra of isolated and paired gold nanoparticles. The single-particle and dimer responses determine the wavelength- and geometry-dependent pair interaction. **g**, Coupling-to-geometry projection obtained from the calibrated relation between physical gap and effective coupling. The selected gaps $g_1$ and $g_2$ are mapped to the effective couplings $t_1 = t\left(g_1; s = \frac{D}{2}\right)$ and $t_2 = t\left(g_2; s = \frac{D}{2}\right)$, enabling inverse design directly in geometry space. **h**, Full-wave validation of the

selected finite plasmonic lattice with $D = 300\ \mathrm{nm}$, $g_1 = 90\ \mathrm{nm}$ and $g_2 = 10\ \mathrm{nm}$, showing a boundary-concentrated electric-field response at $\lambda = 642\ \mathrm{nm}$.

## Equilibrium sensing of nonlocal boundary response

The trained equilibrium sensor resolves a finite-system boundary-response domain that departs from the ideal nearest-neighbor SSH boundary as nonlocal coupling increases. Each sample is specified by the coupling triplet $(t_1, t_2, t_3)$, where $t_1$ and $t_2$ denote the intracell and intercell couplings and $t_3$ represents the longer-range off-diagonal coupling (Fig. 2a). The sensor maps the corresponding nonlocal Hamiltonian onto continuous response scores. A trainable sensor couples to selected lattice regions (Fig. 2b): in one dimension, its outputs are grouped into reference-topological and reference-trivial scores, whereas in two dimensions they are grouped into boundary-localized and trivial scores resolved through corner, edge and bulk sectors.

Training uses symmetric finite-nudging quantum equilibrium propagation. For each coupling triplet, the free equilibrium output is compared with the target, and the resulting error is applied as a weak perturbation to the sensor observables. The collective response of the trainable Hamiltonian terms then updates the sensor parameters. Thus, the SSH couplings define the input, the sensor and system-sensor interactions form the trainable sector, and the outputs remain equilibrium probabilities. The resulting model is a Hamiltonian-native sensor rather than an external regressor acting on post-processed spectra or field maps. The complete Hamiltonian, output projectors, gradient estimator, loss function and optimization protocol are given in Methods.

We first evaluate the framework on a one-dimensional nonlocal SSH chain with a known nearest-neighbor phase boundary. In the limit, $t_3 = 0$, $|t_2| > |t_1|$ and $|t_1| > |t_2|$ define the topological and trivial reference sectors, respectively.[9] Samples close to the clean gap closing are excluded by a finite margin. During training, however, the complete nonlocal Hamiltonian is supplied to the sensor, testing whether this reference distinction remains identifiable when $t_3$ modifies the finite-chain spectrum and spatial response.[23,24] A single sensor qubit couples to the left-edge, right-edge and bulk regions and maps each triplet onto continuous reference-topological and reference-trivial scores. We then extend the task to a finite two-dimensional SSH lattice with dimerized intracell and intercell couplings and nonlocal hopping along both directions. Open boundaries permit direct resolution of edge- and corner-localized states. The lattice is partitioned into corner, edge and bulk sectors, to which a two-qubit sensor is coupled (Fig. 2b); its four outcomes are grouped into boundary-localized and trivial response scores. The training labels are obtained from an operational finite-size diagnostic. For each coupling triplet, the bare lattice Hamiltonian is diagonalized and states within a prescribed midgap window are tested for their boundary weight. A sample is assigned to the boundary-localized class when at least one midgap state exceeds the prescribed threshold, while trivial and ambiguous samples are assigned according to the criteria detailed in Methods. This device-oriented diagnostic is not a substitute for a bulk invariant, but a finite-lattice label for the boundary responses relevant to plasmonic structures. The corresponding spectrum and representative corner- and edge-localized eigenstates are shown in Supplementary Fig. S2. Fig. 2c shows the one-dimensional reference-topological response in the $(t_1, t_2)$ plane after averaging over sampled values of $t_3$. The dashed line marks the clean nearest-neighbor condition $t_1 = t_2$. The learned response departs systematically from this boundary, showing that the finite-system response cannot be inferred from the nearest-neighbor dimerization rule alone. Representative coupling pairs are examined in Fig. 2d, where the sensor response is plotted as a function of $t_3$. Configurations that remain above the classification threshold retain a stable learned response, whereas curves that approach or cross the threshold are sensitive to longer-range coupling. The class-resolved score distributions and initial grouped-output confusion matrix are shown in Supplementary Fig. S1. A useful configuration must therefore lie sufficiently deep within the learned response domain to remain identifiable as nonlocal coupling varies. Its physical topological classification is established in the next section using the nonlocal winding number and a finite-gap condition.

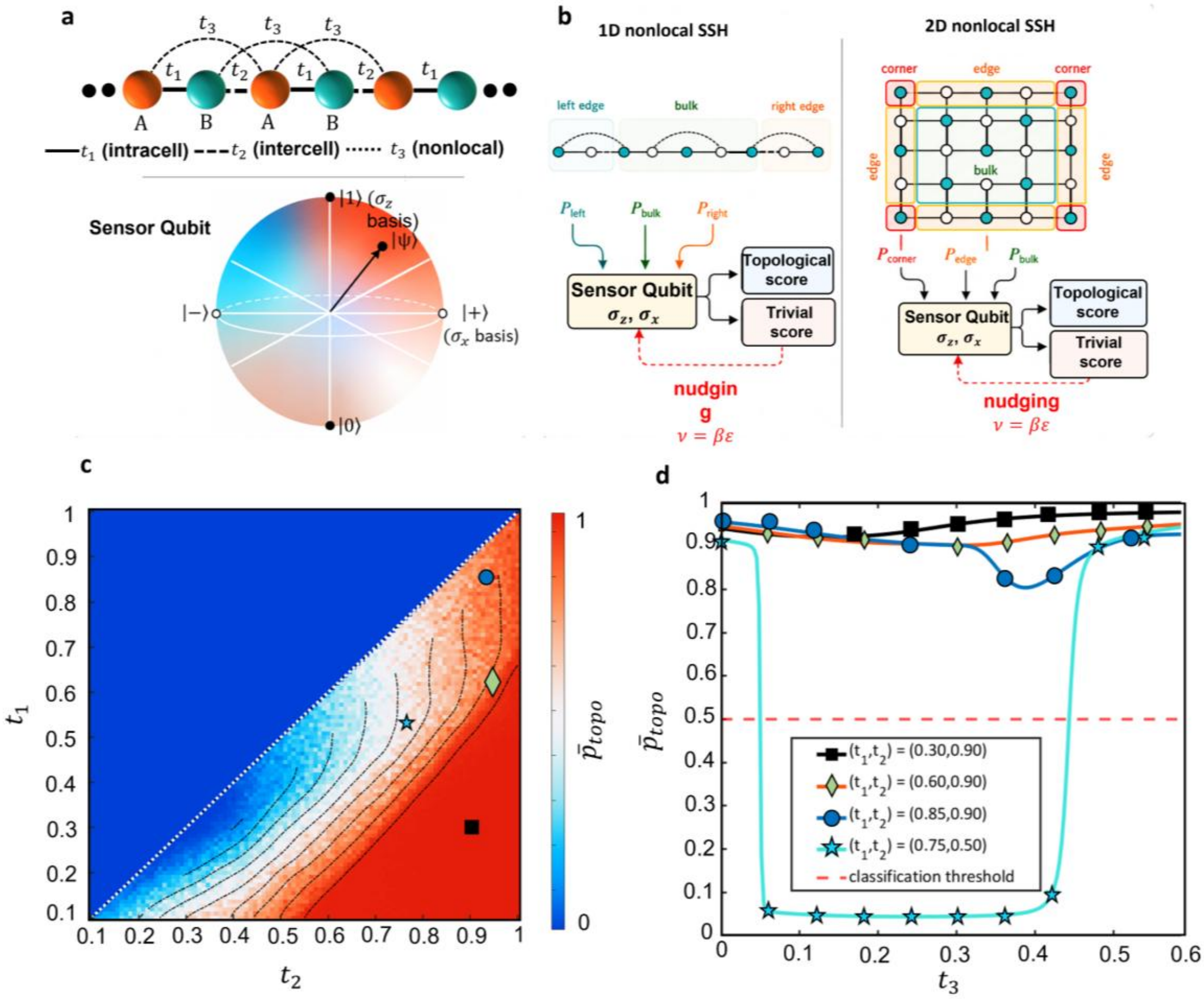

**Fig. 2 | Equilibrium sensing of nonlocal boundary response. a,** Nonlocal one-dimensional SSH lattice used as the controlled benchmark. The intracell and intercell couplings are denoted by $t_1$ and $t_2$, while $t_3$ introduces a longer-range off-diagonal coupling that perturbs the nearest-neighbour SSH limit. The physical lattice is coupled to a sensor qubit whose equilibrium state is read out in the $\sigma_z$ and $\sigma_x$ bases. **b**, Spatially resolved QEP sensing scheme for one- and two-dimensional lattices. In the one-dimensional chain, the sensor couples separately to left-edge, bulk and right-edge sectors. In the finite two-dimensional lattice, the sensor resolves corner, edge and bulk regions, allowing the output to distinguish boundary-localized and trivial responses. The output error is applied through finite nudging during training. **c**, Nonlocality-averaged QEP response map in the $(t_1, t_2)$ plane. The color scale shows the averaged topological or boundary-response score $\bar{p}_{topo}$, obtained after averaging over sampled $t_3$. The dashed diagonal marks the clean nearest-neighbor condition $t_1 = t_2$. The learned response deviates from the ideal SSH boundary, indicating sensitivity to nonlocal coupling. **d**, Response curves for representative coupling pairs marked in **c**, showing the evolution of $\bar{p}_{topo}$ as $t_3$ is varied. Configurations that remain above the classification threshold retain a stable learned boundary response under nonlocal perturbation, whereas curves that cross the threshold indicate nonlocality-sensitive phase assignment.

## Physical relabeling of the learned boundary response

The QEP sensors learn the response associated with the initial sampler labels, which provide a controlled training reference but cannot serve as final topology labels once longer-range hopping is included. In the nearest-neighbor SSH model, the phase boundary is determined by the relative magnitudes of $t_1$ and $t_2$. Nonlocal hopping displaces both the winding structure and the gap closing from this reference boundary, so a high QEP edge- or corner-response score cannot by itself establish a topological phase. We therefore apply physics-based relabeling after QEP training and before inverse design. For the off-diagonal nonlocal SSH model, the Bloch function is

$$q_{nl}(k) = t_1 + t_2 e^{-ik} + t_3 e^{-2ik} \qquad (1)$$

where $t_3$ denotes a longer-range hopping between opposite sublattices, preserving the chiral off-diagonal structure required for winding-number classification.[21,24,25] The nonlocal winding number is

$$\nu = \frac{1}{2\pi}\int_{-\pi}^{\pi}\frac{\partial}{\partial k} arg[q_{nl}(k)]dk \qquad (2)$$

and the corresponding Bloch gap is

$$\triangle_{gap} = 2\min|q_{nl}(k)| \qquad (3)$$

For the symmetric two-dimensional lattice, the winding test is applied along both lattice directions. A sample is assigned to the physically topological class only when

$$\nu_x \geq 1, \nu_y \geq 1,\ \triangle_{gap} > \triangle_{min}. \qquad (4)$$

where $\triangle_{min}$ is a finite gap threshold that excludes samples whose phase assignment is sensitive to numerical, finite-size or fabrication-scale perturbations. Samples with $\triangle_{gap} \leq \triangle_{min}$ are classified as gap ambiguous and are excluded from the robust-design domain.

Fig. 3 shows the effect of this physical relabeling. The winding classification separates topological and trivial regions, while the nonlocal Bloch gap identifies configurations close to gap closure (Fig. 3a,b). Combining both conditions yields the physically relabeled classes in Fig. 3c. The trained QEP sensor assigns high edge- or corner-response scores predominantly to the physically topological region and low scores to most trivial configurations (Fig. 3d). The resulting confusion matrix gives a sensitivity of 99.8% for physically topological samples and a specificity of 98.1% for physically trivial samples (Fig. 3e). The complete score distributions and physically classified coupling-space map are shown in Supplementary Fig. S3. Gap-ambiguous samples provide the clearest motivation for the topology gate. Although some receive high QEP edge- or corner-response scores, their small Bloch gaps make the physical classification unstable. The QEP output is therefore interpreted as a continuous finite-system boundary-response measure rather than as the definition of topology. The winding and gap conditions determine the physical class independently, while the trained sensor quantifies how strongly that verified topology is expressed in the finite lattice.

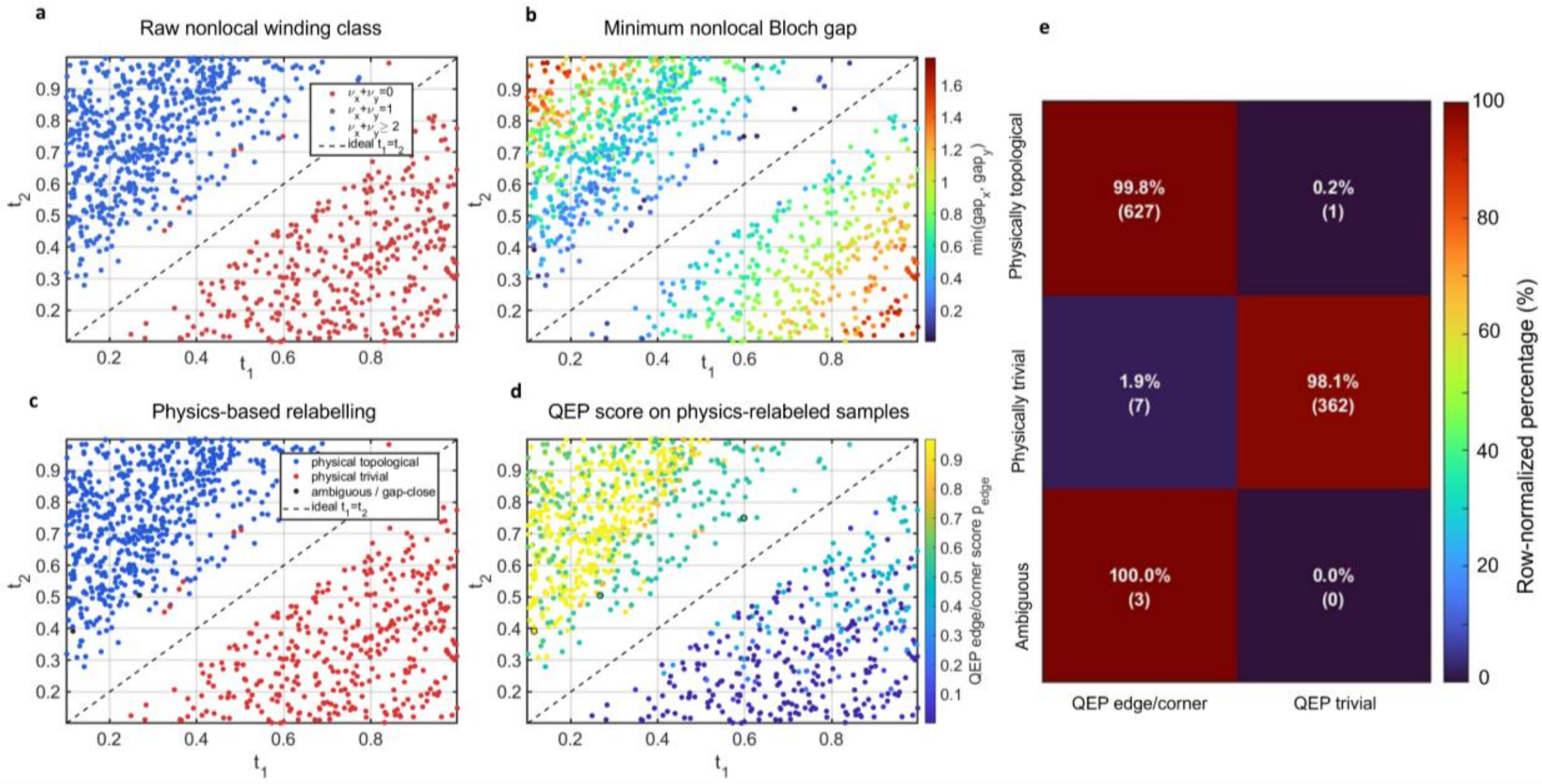

**Fig. 3 | Physical relabeling of the QEP response in nonlocal coupling space. a**, Raw nonlocal winding classification of sampled coupling configurations in the $(t_1, t_2)$ plane. **b**, Minimum nonlocal Bloch gap for the same samples, identifying configurations close to gap closure. **c**, Physics-based relabeling obtained by combining the directional winding condition with a finite-gap threshold. Samples are assigned as physically topological, physically trivial or gap-ambiguous. **d**, QEP edge- or corner-response score evaluated on the physically relabeled samples. **e**, Row-normalized confusion matrix comparing the trained QEP response assignment with the independently determined physical classes. The sensor identifies physically topological samples with 99.8% sensitivity and physically trivial samples with 98.1% specificity. All three gap-ambiguous samples receive an edge/corner QEP assignment, demonstrating why the independent topology gate is required; these samples are subsequently rejected because they do not satisfy the finite-gap condition.

## Physics-gated robustness ranking

After relabeling, we define a conservative topology membership $\mu_{\text{top}}$, which equals one only for winding-verified and gap-open samples. The final design reliability score is

$$\mu_{\text{design}} = \mu_{\text{top}}\, F\big(\mu_{\text{QEP}}, \mu_{\text{contrast}}, \mu_{\text{gap}}, \mu_{\text{NL}}\big) \qquad (5)$$

Here $F$ is the fuzzy aggregation rule. The quantities $\mu_{\text{QEP}}, \mu_{\text{contrast}}, \mu_{\text{gap}}$, and $\mu_{\text{NL}}$ quantify the boundary response, output contrast, gap stability and compatibility with the selected nonlocality regime, respectively. Their normalization ranges and aggregation weights are given in Methods. The multiplicative factor $\mu_{\text{top}}$ ensures that candidates with an invalid winding number or insufficient gap receive $\mu_{\text{design}} = 0$, even when the sensor assigns a high boundary-response score. The fuzzy layer therefore does not define topology, but ranks only configurations that have passed the winding and gap conditions.

The effect of this physically gated ranking is summarized in Fig. 4. The design-decision matrix in Fig. 4a shows that all 369 physically trivial samples and all three gap-ambiguous samples are rejected. Among the physically topological samples, 72.6% (456 samples) are retained as robust designs, whereas 27.4% (172 samples) are rejected because their combined reliability score remains below the acceptance threshold. Fig. 4b locates these groups in the $(t_1, t_2)$ coupling plane, distinguishing accepted robust topological designs from rejected trivial and weak or ambiguous configurations. Fig. 4c shows how the final reliability score varies jointly with the effective Bloch gap and the nonlocality measure $\eta_{\mathrm{NL}}$, which quantifies the relative strength of the longer-range coupling with the color scale representing $\mu_{\mathrm{design}}$. Together, these results show that winding verification is necessary but not sufficient for design selection: only physically topological configurations with a sufficiently strong and stable finite-system boundary response are retained for inverse design.

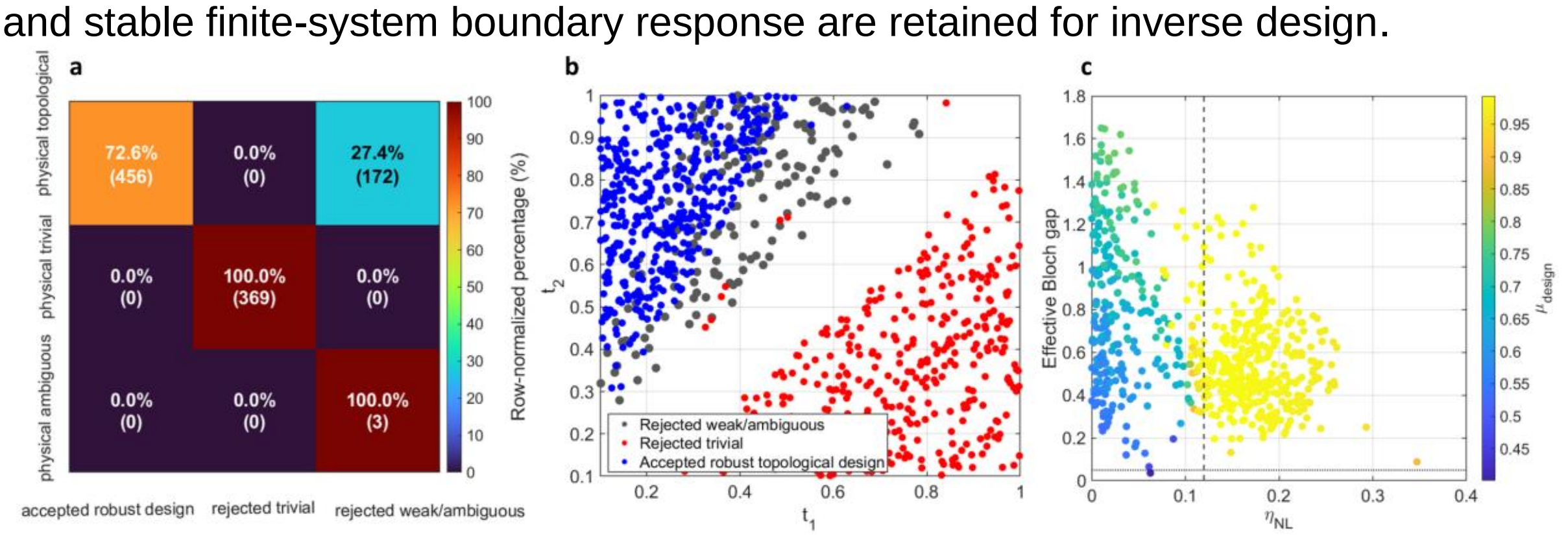

**Fig 4 | Physics-gated robustness ranking after relabeling**. **a**, Physics-gated design-decision matrix showing the final design decision after applying the topology gate and robustness ranking. Physically trivial and gap-ambiguous samples are rejected, while physically topological samples are separated into accepted robust designs and rejected weak or ambiguous candidates. **b**, Distribution of accepted and rejected configurations in the $(t_1, t_2)$ plane. **c**, Dependence of the final design reliability score ($\mu_{\mathrm{design}}$) on the effective Bloch gap and the nonlocality measure $\eta_{\mathrm{NL}}$. The dashed line marks the reference nonlocality value used in the robustness score. Only configurations with verified topology, sufficient gap and stable boundary response are retained for inverse design.

The physics-gated design decision is insensitive to moderate variations of the fuzzy aggregation weights. When each component weight is perturbed by up to 0.05, the accepted design set retains a median Jaccard overlap of 0.983, with a 5th-95th percentile interval of 0.950-0.998. The highest-ranked candidate remains accepted for all tested weight combinations. The full perturbation statistics and candidate-level acceptance map are provided in Supplementary Fig. S4 and Supplementary Table S1.

## Full-wave geometry-to-coupling calibration and inverse design

The preceding analysis yields a physically verified robustness landscape in effective-coupling space.

$$(t_1, t_2, t_3) \longmapsto \mu_{design} \qquad (6)$$

This landscape identifies configurations that combine a valid nonlocal winding number, a strong QEP boundary response, sufficient gap stability and compatibility with the selected nonlocality regime. To convert this coupling-space result into a physical design rule, we construct a full-wave geometry-to-coupling calibration using COMSOL simulations of isolated and paired gold nanoparticles. The same calibration could be obtained experimentally from measured spectra, thereby incorporating fabrication imperfections and geometry-dependent disorder directly into the geometry-to-coupling map. For a nanoparticle pair with radius $s$ and gap $g$, the dimer extinction spectrum $S_{dimer}(\lambda; g, s)$ is compared with the isolated-particle spectrum $S_{single}(\lambda; s)$. The single-particle spectrum defines the geometry-dependent resonance of the building block, whereas the dimer spectrum contains interaction-induced resonance shifts, oscillator-strength redistribution and mode splitting associated with electromagnetic hybridization (Fig. 5a,b).[39-41] These spectral changes are reduced to an effective wavelength-resolved pair coupling $t_{pair}(g, s; \lambda)$, as detailed in Methods. The calibrated coupling decreases with increasing gap, distinguishing the strongly hybridized regime at small separations from the weakly coupled regime at larger separations (Fig. 5c). A candidate geometry is specified by the particle radius s and the intracell and intercell gaps $\boldsymbol{g} = (g_1, g_2)$, which map the physical structure onto the effective-coupling representation used by the inverse-design objective:

$$\mathbf{t}_{eff}(s, \mathbf{g}; \lambda) = [t_1, t_2] = \left[t_{pair}(g_1, s; \lambda), t_{pair}(g_2, s; \lambda)\right] \qquad (7)$$

The QEP sensor remains a coupling-space evaluator, while the calibration layer supplies the physical coordinates required for geometry-level inverse design. Each calibrated pair $(t_1, t_2)$ is evaluated together with $t_3$ over the selected nonlocality range using the same QEP boundary response, nonlocal winding number, Bloch gap and design-reliability criteria.

At the geometry level, the inverse-design objective is

$$J(s, g_1, g_2) = -\min_{t_3 \in \tau_{\mathrm{NL}}} \mu_{\mathrm{design}}\left[\mathbf{t}_{eff}(s, g_1, g_2), t_3\right] + \lambda_{\mathrm{pen}} \mathcal{P}(s, g_1, g_2, t_3) \qquad (8)$$

where $\mu_{\mathrm{design}}$ is the winding-gated robustness score and $\mathcal{P}$ penalizes fabrication-incompatible gaps, undesired coupling contrasts or geometries outside the calibrated range. The selected geometry is

$$(s^*, g_1^*, g_2^*) = arg \min_{s, g_1, g_2} J(s, g_1, g_2) \qquad (9)$$

Minimizing $J$ therefore maximizes the physics-gated design reliability while satisfying the fabrication constraints. The extinction-cross-section calibration was generated for isolated particles and dimers over a nanoparticle-radius range of $s = 20 - 250\ \mathrm{nm}$ and the wavelength range of $300 - 1000\ \mathrm{nm}$ (The corresponding radius-resolved dimer extinction maps are shown in Supplementary Fig. S5). For the inverse-design demonstration, we fix the target operating wavelength at $\lambda = 650\ \mathrm{nm}$, where a boundary-localized response is desired, and restrict the particle radius to $s = 120 - 160\ \mathrm{nm}$. Within this design space, the highest-ranked geometry is obtained for $s^* = 150\ \mathrm{nm}$, $g_1^* = 90\ \mathrm{nm}$ and $g_2^* = 10\ \mathrm{nm}$. The complete geometry-space objective landscape and the partition of the selected finite lattice into corner, edge and bulk sectors are shown in Supplementary Fig. S6. The smaller intercell gap $g_2$ lies in the strongly hybridized regime, whereas the larger intracell gap $g_1$ lies in the weakly coupled regime. The resulting hierarchy $t_2 > t_1 > t_3$ is evaluated using the nonlocal winding and finite-gap conditions and lies within the verified topological design domain. Full-wave simulations of the selected lattice are shown in Fig. 5d-f. The inverse-designed geometry supports a boundary-concentrated field response in the target spectral region, whereas off-resonant wavelengths redistribute the field into distinct spatial patterns. Because the complete electromagnetic lattice is simulated directly, these calculations intrinsically include interactions beyond the calibrated nearest-neighbor bonds.

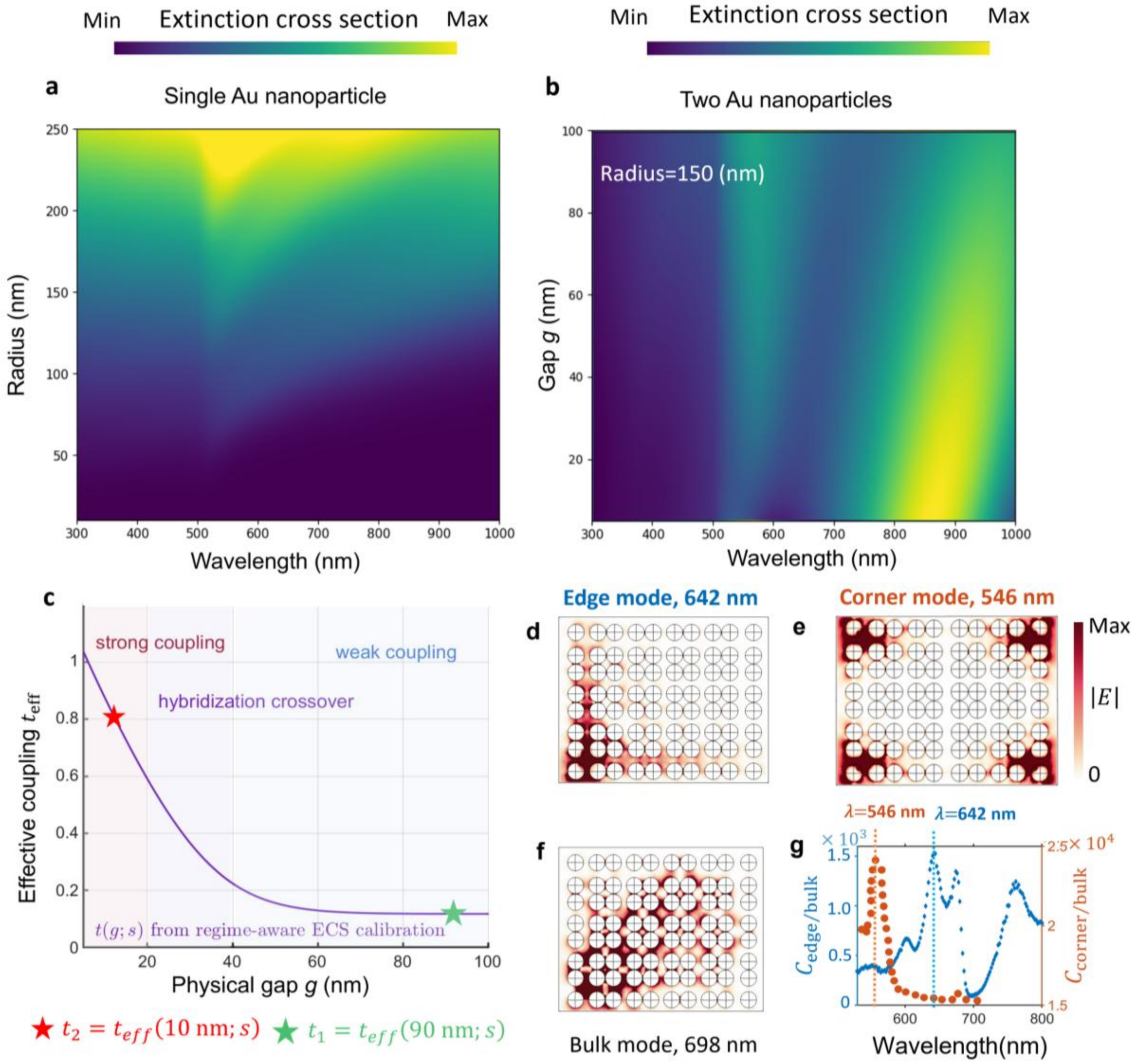

**Fig. 5 | Full-wave geometry-to-coupling calibration and inverse-designed plasmonic lattice. a**, Full-wave extinction response of an isolated gold nanoparticle as a function of wavelength and particle radius. The single-particle spectrum defines the geometry-dependent optical resonance of the building block. **b**, Extinction response of a gold nanoparticle dimer with particle radius $s = 150\ \mathrm{nm}$ plotted as a function of wavelength and gap $g$. The spectral shifts, redistribution and splitting relative to the isolated-particle response arise from electromagnetic hybridization. **c**, Regime-aware geometry-to-coupling mapping for the selected particle radius and design wavelength. The normalized effective pair coupling $t_{pair}(g, s; \lambda)$ extracted from the full-wave extinction calibration. Smaller gaps correspond to stronger hybridization, while larger gaps approach the weak-coupling regime. The shaded region indicates the crossover identified from changes in the dimer extinction spectra. The calibrated gaps $g_1$ and $g_2$ define the intracell and intercell SSH couplings used in the inverse-designed lattice. The inverse-design search was restricted to $\lambda = 650\ \mathrm{nm}$ and particle radii $s = 120 - 160\ \mathrm{nm}$. The selected geometry has $s^* = 150\ \mathrm{nm}$, $g_1^* = 90\ \mathrm{nm}$ and $g_2^* = 10\ \mathrm{nm}$. **d–f**, Full-wave electric-field distributions of the selected finite plasmonic lattice. The field response is edge dominated at 642 nm (**d**), corner dominated at 546 nm (**e**) and bulk dominated at 698 nm (**f**). These calculations validate that the geometry selected from the physics-gated coupling-space landscape supports spectrally distinct finite-lattice spatial responses. **g**, Wavelength-dependent edge-to-bulk and corner-to-bulk field contrasts obtained from sector-resolved sampling of the full-wave fields. The edge-to-bulk contrast exhibits a characteristic maximum near 642 nm, whereas the corner-to-bulk contrast peaks near 546 nm, consistent with the field distributions in **d** and **e**, respectively.

To quantify full-wave localization, identical sampling volumes are defined around each nanoparticle. The boundary sector contains the corner and edge particles, whereas the bulk sector contains the interior particles. The wavelength-dependent boundary-to-bulk intensity contrast is

$$C_{\partial/\mathrm{bulk}}(\lambda) = \frac{V_{\partial}^{-1} \int_{V_{\partial}} |E(\mathbf{r},\lambda)|^2 dV}{V_{\mathrm{bulk}}^{-1} \int_{V_{\mathrm{bulk}}} |E(\mathbf{r},\lambda)|^2 dV} \quad (10)$$

Corner-to-bulk and edge-to-bulk contrasts are defined analogously by replacing $V_{\partial}$ with the corresponding corner or edge sampling region. The resulting spectra show distinct boundary-enhanced resonances associated with the corner- and edge-dominated responses (Fig. 5g). Comparison with the full-wave field maps identifies 546 nm as corner dominated and 642 nm as edge dominated, with corresponding sector-to-bulk contrasts of $2.4 \times 10^4$ and $1.5 \times 10^3$, respectively (Fig. 5d-f). The present calibration is pairwise, with each effective bond inferred from the extinction response of an isolated particle and a dimer. This approximation keeps the geometry-to-Hamiltonian map interpretable and transferable, while defining its regime of validity. Dense or strongly hybridized arrays may require cluster-level calibration that explicitly includes the neighboring geometry. The same calibration principle can be extended to nanodisks, nanoholes, dielectric resonators, hybrid plasmonic-excitonic,[42] van der Waals[43] and photonic-magnonic platforms[44-48] by recalculating the corresponding isolated-element and pair or cluster responses. In each case, the QEP sensor remains a coupling-space evaluator, while the calibration layer is adapted to the physical platform.

## Discussion

Here we show that nonlocality can be treated as a design variable rather than as an uncontrolled deviation from the ideal SSH limit. In realistic plasmonic lattices, long-range electromagnetic coupling is unavoidable: it is the mechanism that links the building blocks, but it also reshapes the phase structure inherited from short-range Hamiltonians. This dual role creates a gap between topological classification in effective models and topological design in physical geometry. The framework developed here bridges this gap by separating the design problem into distinct physical operations: equilibrium sensing in coupling space, winding-based relabeling, robustness ranking, and full-wave geometry-to-coupling calibration. Each layer has a separate role. The learned QEP response does not define topology; the winding number and gap condition verify it. The fuzzy layer does not create a phase boundary; it ranks physically verified candidates. The calibration layer does not train the sensor; it projects realistic geometries onto the learned and physically corrected coupling-space landscape.

This physically ordered architecture provides a physics-informed strategy for machine learning in topological photonics. Rather than training a black-box surrogate to assign a phase label directly from spectra or geometry, the trainable component remains embedded in a Hamiltonian description. Its inputs are effective couplings, its trainable degrees of freedom are sensor couplings, and its outputs are equilibrium observables associated with finite-system boundary response. This ordering is particularly important for finite plasmonic lattices, where device function is governed not only by a bulk invariant but also by the persistence of measurable edge- or corner-localized responses under nonlocal coupling, spectral dispersion and fabrication-scale perturbations. The full-wave sector-resolved contrasts further distinguish corner-,

edge- and bulk-dominated responses, showing how the independently verified topology is expressed in the physical lattice. Several limitations define the present regime of validity. The finite-lattice boundary diagnostic is device-oriented and should not be interpreted as a universal substitute for bulk topological invariants. The pairwise calibration captures the dominant gap-dependent interaction between two building blocks, but dense or strongly hybridized arrays may require cluster-level calibration in which the effective coupling depends explicitly on neighboring geometry. For metallic plasmonic systems, dissipative and non-Hermitian effects may also need to be incorporated explicitly because linewidth, loss and mode hybridization change both spectral calibration and boundary-mode observability. These limitations do not alter the central design principle established here. Robust topological inverse design should therefore rely neither on the nearest-neighbor SSH rule alone nor on an unconstrained classifier. Instead, topology must be verified independently, while finite-system response, robustness and physical realizability are evaluated as separate design quantities.
More broadly, the calibration layer is not restricted to passive plasmonic particles. It can be extended to magneto-optic, magnetoplasmonic and photonic-magnonic heterostructures in which an applied field, magnetic-domain configuration or magnonic mode provides an additional reconfigurable control coordinate.[27-29,47] The corresponding map $\mathbf{t}_{eff}(s, \mathbf{g}, \lambda, \mathbf{H}, \mathbf{m})$ can be obtained by combining full-wave electromagnetic simulations, micromagnetic modeling and experiments, thereby linking the magnetic state and magnonic mode structure to the effective intracell, intercell and nonlocal couplings.[30,31,46,47] Magnetic domain walls and patterned magnetic configurations could then provide mechanisms for switching, displacing or spatially reshaping photonic boundary responses without reconstructing the underlying lattice.[28,29,31,44,45] Low-damping magnetic materials supporting coherent magnon transport and strong magnon–photon coupling could further extend this strategy towards quantum-magnonic devices.[47,48] This extension moves geometry-calibrated topological design from passive optimization towards actively reconfigurable photonic-magnetic and photonic-magnonic heterostructures.

## References

1. Su, W. P., Schrieffer, J. R. & Heeger, A. J. Solitons in polyacetylene. *Phys. Rev. Lett*. **42**, 1698-1701 (1979). DOI: 10.1103/PhysRevLett.42.1698.

2. Haldane, F. D. M. & Raghu, S. Possible realization of directional optical waveguides in photonic crystals with broken time-reversal symmetry. *Phys. Rev. Lett*. **100**, 013904 (2008). DOI: 10.1103/PhysRevLett.100.013904.

3. Wang, Z., Chong, Y. D., Joannopoulos, J. D. & Soljačić, M. Observation of unidirectional backscattering-immune topological electromagnetic states. *Nature* **461**, 772-775 (2009). DOI: 10.1038/nature08293.

4. Hafezi, M., Mittal, S., Fan, J., Migdall, A. & Taylor, J. M. Imaging topological edge states in silicon photonics. *Nat. Photon*. **7**, 1001-1005 (2013). DOI: 10.1038/nphoton.2013.274.

5. Lu, L., Joannopoulos, J. D. & Soljačić, M. Topological photonics. *Nat. Photon*. **8**, 821-829 (2014). DOI: 10.1038/nphoton.2014.248.

6. Khanikaev, A. B. & Shvets, G. Two-dimensional topological photonics. *Nat. Photon*. **11**, 763-773 (2017). DOI: 10.1038/s41566-017-0048-5.

7. Ozawa, T. *et al*. Topological photonics. *Rev. Mod. Phys*. **91**, 015006 (2019). DOI: 10.1103/RevModPhys.91.015006.

8. Kim, M., Jacob, Z. & Rho, J. Recent advances in 2D, 3D and higher-order topological photonics. *Light Sci. Appl*. **9**, 130 (2020). DOI: 10.1038/s41377-020-0331-y.

9. Khanikaev, A. B. & Alù, A. Topological photonics: robustness and beyond. *Nat. Commun*. **15**, 931 (2024). DOI: 10.1038/s41467-024-45194-2.

10. Poli, C., Bellec, M., Kuhl, U., Mortessagne, F. & Schomerus, H. Selective enhancement of topologically induced interface states in a dielectric resonator chain. *Nat. Commun*. **6**, 6710 (2015). DOI: 10.1038/ncomms7710.

11. Noh, J. *et al*. Topological protection of photonic mid-gap defect modes. *Nat. Photon*. **12**, 408-415 (2018). DOI: 10.1038/s41566-018-0179-3.

12. Benalcazar, W. A., Bernevig, B. A. & Hughes, T. L. Quantized electric multipole insulators. *Science* **357**, 61-66 (2017). DOI: 10.1126/science.aah6442.

13. Mittal, S. *et al*. Photonic quadrupole topological phases. *Nat. Photon*. **13**, 692-696 (2019). DOI: 10.1038/s41566-019-0452-0.

14. El Hassan, A. *et al*. Corner states of light in photonic waveguides. *Nat. Photon*. **13**, 697-700 (2019). DOI: 10.1038/s41566-019-0519-y.

15. Li, M. *et al*. Higher-order topological states in photonic kagome crystals with long-range interactions. *Nat. Photon*. **14**, 89-94 (2020). DOI: 10.1038/s41566-019-0561-9.

16. He, L. *et al*. Quadrupole topological photonic crystals. *Nat. Commun*. **11**, 3119 (2020). DOI: 10.1038/s41467-020-16916-z.

17. Xie, B. *et al*. Higher-order quantum spin Hall effect in a photonic crystal. *Nat. Commun*. **11**, 3768 (2020). DOI: 10.1038/s41467-020-17593-8.

18. Kim, H.-R. *et al*. Multipolar lasing modes from topological corner states. *Nat. Commun*. **11**, 5758 (2020). DOI: 10.1038/s41467-020-19609-9.

19. Davoodi, F. From bound states to quantum spin models: chiral coherent dynamics in topological photonic rings. *Nanophotonics* **14**, 4397-4409 (2025). DOI: 10.1515/nanoph-2025-0473.

20. Davoodi, F. Controlling collective quasiparticle dynamics beyond decoherence in topological interfaces. *Laser Photon. Rev*. **e71152** (2026). DOI: 10.1002/lpor.71152.

21. Kim, G., Suh, J., Lee, D., Park, N. & Yu, S. Long-range-interacting topological photonic lattices breaking channel-bandwidth limit. *Light Sci. Appl*. **13**, 189 (2024). DOI: 10.1038/s41377-024-01557-4.

22. Roccati, F. *et al*. Hermitian and non-Hermitian topology from photon-mediated interactions. *Nat. Commun*. **15**, 2400 (2024). DOI: 10.1038/s41467-024-46471-w.

23. Pocock, S. R., Xiao, X., Huidobro, P. A. & Giannini, V. Topological plasmonic chain with retardation and radiative effects. *ACS Photon*. **5**, 2271–2279 (2018). DOI: 10.1021/acsphotonics.8b00117.

24. Downing, C. A. & Weick, G. Topological plasmons in dimerized chains of nanoparticles: robustness against long-range quasistatic interactions and retardation effects. *Eur. Phys. J. B* **91**, 253 (2018). DOI: 10.1140/epjb/e2018-90199-0.

25. Rosiek, C. A. *et al*. Observation of strong backscattering in valley-Hall photonic topological interface modes. *Nat. Photon*. **17**, 386-392 (2023). DOI: 10.1038/s41566-023-01189-x.

26. Gao, F. *et al*. Probing topological protection using a designer surface plasmon structure. *Nat. Commun*. **7**, 11619 (2016). DOI: 10.1038/ncomms11619.

27. Jin, D. *et al*. Topological magnetoplasmon. *Nat. Commun*. **7**, 13486 (2016). DOI: 10.1038/ncomms13486.

28. Jin, D. *et al*. Topological kink plasmons on magnetic-domain boundaries. *Nat. Commun*. **10**, 4565 (2019). DOI: 10.1038/s41467-019-12092-x.

29. Tang, W., Wang, M., Ma, S., Chan, C. T. & Zhang, S. Magnetically controllable multimode interference in topological photonic crystals. *Light Sci. Appl*. **13**, 112 (2024). DOI: 10.1038/s41377-024-01433-1.

30. McCord, J. Progress in magnetic domain observation by advanced magneto-optical microscopy. *J. Phys. D Appl. Phys*. **48**, 333001 (2015). DOI: 10.1088/0022-3727/48/33/333001.

31. Trützschler, J., Sentosun, K., Mozooni, B., Mattheis, R. & McCord, J. Magnetic domain wall gratings for magnetization reversal tuning and confined dynamic mode localization. *Sci. Rep*. **6**, 30761 (2016). DOI: 10.1038/srep30761.

32. Molesky, S. *et al*. Inverse design in nanophotonics. *Nat. Photon*. **12**, 659-670 (2018). DOI: 10.1038/s41566-018-0246-9.

33. Melati, D. *et al*. Mapping the global design space of nanophotonic components using machine learning pattern recognition. *Nat*. *Commun*. **10**, 4775 (2019). DOI: 10.1038/s41467-019-12698-1.

34. Peurifoy, J. *et al*. Nanophotonic particle simulation and inverse design using artificial neural networks. *Sci. Adv*. **4**, eaar4206 (2018). DOI: 10.1126/sciadv.aar4206.

35. Davoodi, F. Active physics-informed deep learning: surrogate modeling for nonplanar wavefront excitation of topological nanophotonic devices. *Nano Lett*. **25**, 768-775 (2025). DOI: 10.1021/acs.nanolett.4c05120.

36. Wanjura, C. C. & Marquardt, F. Quantum equilibrium propagation for efficient training of quantum systems based on Onsager reciprocity. *Nat. Commun*. **16**, 6595 (2025). DOI: 10.1038/s41467-025-61665-6.

37. Laydevant, J., Marković, D. & Grollier, J. Training an Ising machine with equilibrium propagation. *Nat. Commun*. **15**, 3671 (2024). DOI: 10.1038/s41467-024-46879-4.

38. Scellier, B. & Bengio, Y. Equilibrium propagation: bridging the gap between energy-based models and backpropagation. *Front. Comput. Neurosci*. **11**, 24 (2017). DOI: 10.3389/fncom.2017.00024.

39. Jain, P. K., Huang, W. & El-Sayed, M. A. On the universal scaling behavior of the distance decay of plasmon coupling in metal nanoparticle pairs: a plasmon ruler equation. *Nano Lett*. **7**, 2080-2088 (2007). DOI: 10.1021/nl071008a.

40. Prodan, E., Radloff, C., Halas, N. J. & Nordlander, P. A hybridization model for the plasmon response of complex nanostructures. *Science* **302**, 419-422 (2003). DOI: 10.1126/science.1089171.

41. Nordlander, P., Oubre, C., Prodan, E., Li, K. & Stockman, M. I. Plasmon hybridization in nanoparticle dimers. *Nano Lett*. **4**, 899-903 (2004). DOI: 10.1021/nl049681c.

42. Johnson, P. B. & Christy, R. W. Optical constants of the noble metals. *Phys. Rev. B* **6**, 4370-4379 (1972). DOI: 10.1103/PhysRevB.6.4370.

43. Davoodi, F. & Granpayeh, N. Near-infrared absorbers based on the heterostructures of two-dimensional materials. *Appl. Opt*. **57**, 1358-1366 (2018). DOI: 10.1364/AO.57.001358.

44. Holländer, R. B., Müller, C., Schmalz, J., Gerken, M. & McCord, J. Magnetic domain walls as broadband spin wave and elastic magnetisation wave emitters. *Sci. Rep*. **8**, 13871 (2018). DOI: 10.1038/s41598-018-31689-8.

45. Mozooni, B. & McCord, J. Direct observation of closure domain wall mediated spin waves. *Appl. Phys. Lett*. **107**, 042402 (2015). DOI: 10.1063/1.4927598.

46. Fabiha, R. *et al*. Spin wave electromagnetic nano-antenna enabled by tripartite phonon–magnon-photon coupling. *Adv. Sci*. **9**, 2104644 (2022). DOI: 10.1002/advs.202104644.

47. Flebus, B. *et al*. The 2024 magnonics roadmap. *J. Phys.: Condens. Matter* **36**, 363501 (2024). DOI: 10.1088/1361-648X/ad399c.

48. Serha, R. O., Dubs, C. & Chumak, A. V. Magnetic materials for quantum magnonics. *APL Mater*. **14**, 030901 (2026). DOI: 10.1063/5.0306423.